# Chemical, Structural, and Transport Properties of $Na_{1-x}CoO_2$

F. Rivadulla, J.-S. Zhou, J. B. Goodenough
Texas Materials Institute, ETC 9.102, The University of Texas at Austin
1 University Station, C2201, Austin, TX 78712

We report measurement of room-temperature compressibility, thermal expansion, thermoelectric power $\alpha(T)$ at various pressures $P \leq 20$ kbar, basal-plane resistivity $\rho_{ab}$ (T), magnetic susceptibility and thermal conductivity $\kappa(T)$ taken on single-crystal or cold-pressed $Na_{0.57}CoO_2$. An enhancement of a large $\alpha(T)$ with a change of slope occurs on heating near 100 K, but this enhancement is progressively suppressed by pressure $P \leq 20$ kbar. The c-axis thermal expansion is large in the interval 150 K $<T<$ 250 K where the c-axis resistivity $\rho_c(T)$ exhibits a smooth transition from a metallic to a non-metallic temperature dependence; but the basal-plane thermal expansion remains negligible for all temperatures T < 300 K. On the other hand, the basal-plane room-temperature compressibility is large in the interval 0 < P < 22 kbar, becoming negligible in the range 22 < P < 45 kbar, whereas the c-axis room-temperature compressibility is anomalously large in the pressure range 22 < P < 35 kbar. The basal plane resistivity is $\rho_{ab} \propto T^{3/2}$ below 175 K where there is 3D metallic conduction; it rises less rapidly with temperature where the metallic conduction is confined to 2D. The phonon contribution to the thermal conductivity of a cold-pressed ceramic sample is not suppressed, as previously reported. These findings are rationalized with the aid of the virial theorem, recognition of a pinning of the nominal Co(IV)/Co(III) redox couple at the top of the $O^{2-}:2p^6$ bands, and a schematic location of the $a_1^T$ and $e^T$ antibonding bands of this couple with respect to the Fermi energy $E_F$.

## INTRODUCTION

The layered oxides $A_{1-x}CoO_2$ all contain close-packed planes of low-spin, octahedral-site Co atoms sharing common octahedral-site edges; the A atoms occupy octahedral or trigonal-prismatic sites between the layers[1]. The $Li^+$ ions of $Li_{1-x}CoO_2$ remain in octahedral sites for all values of x, and the system has been extensively studied and commercialized since the demonstration[2] that it can be used as the cathode of a $Li^+$-ion rechargeable battery. As Li is extracted at room temperature, a flat open-circuit voltage of a $Li_{1-x}CoO_2$ cell in the range 0.05 < x < 0.25[3] signals a first-order change from polaronic to itinerant in-plane conduction as x increases[4], and oxygen evolution from the $CoO_2$ layers sets in for x > 0.5[5]. The $Na^+$ ions of the system $Na_{1-x}CoO_2$ change from octahedral to trigonal-prismatic coordination across a two-phase region as x increases in the range 0.15 < x< 0.22[6], but the $CoO_2$ sheets remain intact, only becoming displaced relative to one another so long as half or more of the $Li^+$ are retained, *i.e.* x ≤ 0.5. Recent interest in nominal $Na_{0.5}CoO_2$ has been triggered by its potential for thermoelectric cooling[7]; a large thermoelectric power and low in-plane resistivity[8] have been reported to be accompanied by a thermal conductivity as low as that of an amorphous compound[9]. However, the thermal conductivity was measured on a polycrystalline sample of unknown water content; the hygroscopic character of the partially occupied Na layers has recently been highlited by the report of possible superconductivity in highly hydrated nominal $Na_{0.35}CoO_2 \cdot nH_2O$[10]. In this paper, we discuss the chemistry of $Na_{1-x}CoO2 \cdot nH_2O$ and report that the thermal conductivity of a cold-pressed, anhydrous sample of $Na_{0.57}CoO_2$ shows a conventional phonon component. In addition, we report the evolution of the lattice parameters with temperature and pressure showing an anomalous resistance to stretching, but a high compressibility, of the Co-Co separation in the $CoO_2$ planes. We interpret the anisotropictransport properties in the framework of a two-band model, a change from 3D to 2D itinerant-electron conduction with increasing temperature in the range 150 K < T < 200 K, and an in-plane bandwidth that approaches the Mott-Hubbard strong-correlation limit from the itinerant-electron side. The possible appearance of superconductivity in highly hydrated $Na_{0.35}CoO_2 \cdot nH_2O$ is interpreted to follow formation of $Na_{0.35}CoO_2 \cdot 2\delta H_3O^+(n-3\delta)H_2O$.

**SAMPLE CHEMISTRY**

Both single-crystal and polycrystalline ceramic samples were used in this study. The ceramic samples were prepared by the method described by Cushing and Wiley[11]. Stoichiometric amounts of Co metal and anhydrous NaOH were ground under Ar atmosphere and fired at 700°C under flowing oxygen for a total of 5 days with intermediate grindings. This method eliminates the $Co_3O_4$ impurities present after conventional solid-state reaction of $Co_3O_4$ and $Na_2CO_3$. Atomic absorption analysis (AAA) confirmed that the Na content in $Na_xCoO_2$ was always closer to 0.6 than to the nominal 0.5. This result is independent of the method of fabrication; it was also observed for a material synthesized from $Co_3O_4$ and $Na_2CO_3$ in air at 900°C. To reduce the Na content further, chemical or electrochemical extraction is needed. This observation must be applicable to other samples reported on in the literature that were prepared by direct reaction of precursors with Na in excess of stoichiometric proportions, samples that were assumed to be $Na_{0.5}CoO_2$. Moreover, electrochemical or chemical (with $I_2$ or $Br_2$) extraction of Na beyond x≈0.5 was found to be accompanied by a removal of oxygen as has been reported[5] for $Li_xCoO_2$; this situation probably applies also to $K_{1-x}CoO_2$. In addition, the material is moisture-sensitive with waterbecoming inserted into the Na layers and expanding the c-axis once x is reduced to ≈ 0.7. For this reason, we manipulated the specimen with minimum air exposure. In order to get high-density polycrystalline samples, the anhydrous powders were cold-pressed under 80-100 kbar and then annealed at 900°C.

Single crystals (1.5 x 1.5 x 0.01 mm$^3$) were grown from a NaCl flux. $Co_3O_4$, $Na_2CO_3$, and NaCl were mixed in a molar ratio Na:Co:NaCl=1:1:7, placed in an alumina boat and fired at 950°C for 12 h. The temperature was then slowly reduced down to 850°C at 0.5°C/h, and then at 180°C/h to room temperature.

A four-probe method was used to measure the resistivity. Magnetization was measured with a SQUID magnetometer (Quantum Design).Thermal conductivity was measured with a steady-state method; the temperature gradient was controlled to be less than 1% of the base temperature. The powder x-ray diffraction under pressure has been performed in a diamond anvil cell mounted on a four-circle goniometer. The x-ray beam (2.0 kW) is generated from a Mo target in a sealed tube and monochromated by a graphite crystal. NaCl or Au powder mixed with the sample powder and epoxy (no hardener component) is used as the pressure manometer. Powder diffraction peaks show no broadening in pressures up to 5 GPa; they are collected by a Fuji image plate and integrated into a one dimensional I~2θ curve with the program fit2D.

**RESULTS AND DISCUSSION**

Fig. 1 shows the in-plane resistivity $\rho_{ab}(T^{3/2})$ for $Na_{0.57}CoO_2$. In view of a transition from polaronic to itinerant in-plane conduction as x increases beyond x≈0.25 in $Na_{1-x}CoO_2$[12], we have previously[13] attributed the $T^{3/2}$ behavior of $\rho_{ab}(T)$ below 175 K, where the compound exhibits 3D metallic behavior, to the presence of strong-correlation fluctuations persisting into the compositions with x > 0.25. This departure from conventional Fermi-liquid behavior was rationalized within the context of a 3D electron gas, a condition that breaks down progressively above 170 K where the c-axis conductivity becomes thermally activated. The inset of Fig. 1 shows the pressure dependence of the thermoelectric power α(T) of $Na_{0.57}CoO_2$. The large temperature-dependent α(T) has a slope change near 150 K that progressively decreases as the pressure increases.

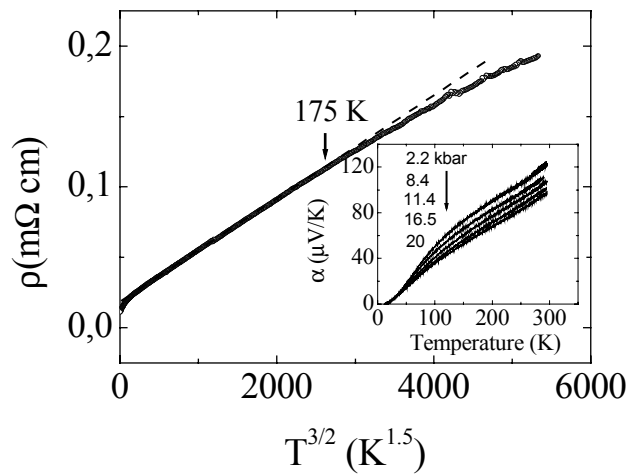

Fig.1: Temperature dependence of the resistivity along the ab-plane versus $T^{3/2}$, and thermoelectric power under pressure of a dense polycrystalline sample (inset).

Fig. 2 shows the evolution of the room-temperature lattice parameters with pressure P < 45 kbar for the powder crushed from single-crystals. Although the structure undergoes no symmetry change in this pressure range, the in-plane and c-axis compressibilities are highly anisotropic with a marked change near 22 kbar. For P < 22 kbar, the in-plane compressibility is large, but it essentially vanishes above 22 kbar whereas the c-axis compressibility increases sharply in the range 22 < P < 35 kbar. The thermal expansion data of Fig. 3 were taken on a polycrystalline sample having a somewhat larger c-axis at room temperature than the single-crystal sample, indicating some difference in the $Na^+$ and/or water content between the two samples. Moreover, the moisture sensitivity of the polycrystalline sample complicated measurement of the thermal expansion, giving relatively broad x-ray-diffraction peaks and introducing some error in the values of a = b and c. Nevertheless, Fig. 3 shows clearly that the in-plane thermal expansion below room temperature is negligible whereas the c-axis thermal expansion is relatively large in the temperature interval 150 K < T < 250 K where the c-axis resistivity changes from itinerant to polaronic with increasing temperature.

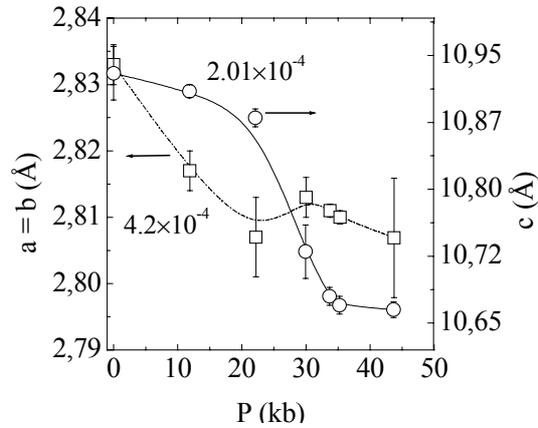

Fig. 2: Pressure dependence of the room temperature lattice parameters obtained in a powder from crushed single crystals. The numbers indicate the compressibility in the ab-plane and along the c-axis, up to 20 kbar. Lines are guides to the eye.

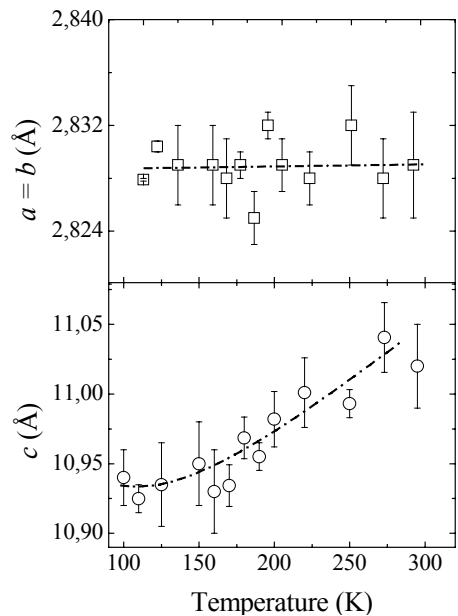

Fig.3: Thermal evolution of the in-plane (above) and along the c-axis (below) lattice parameters of a polycrystalline sample. Si was used as an internal standard. Lines are guides to the eye.

This quite unusual variation of lattice parameters with temperature and pressure complements the transport data and can be understood from the virial theorem, which states that for central-force fields

$$2\langle T\rangle+\langle V\rangle=0$$

where the mean kinetic energy $\langle T\rangle$ of a system of electrons decreases discontinuously if the mean volume of the electrons increases discontinuously across a localized to itinerant electronic transition. Since the electrons are bound, their mean potential energy $\langle V\rangle$ is negative, and a discontinuous decrease in $\langle V\rangle$ for antibonding electrons is accomplished by a discontinuous decrease in the equilibrium M-O bond length. Therefore, crossover occurs at a first-order transition with the localized-electron equilibrium bond length larger than the equilibrium itinerant-electron bond length, *i.e.* (M-O)$_{loc}$ > (M-O)$_{itin}$. It follows that where the two phases coexist, the double-well potential for the equilibrium M-O bond length gives an average <M-O> that is highly compressible and that pressure stabilizes the itinerant-electron phase relative to the localized-electron phase[14]. In order to apply this reasoning to the Na$_{1-x}$CoO$_2$ system, we first recognize that the threefold-degenerate t$_2$ orbitals at the low-spin, octahedral-site cobalt atoms are split by the hexagonal symmetry into an a$_1^T$ orbital directed along the c-axis and twofold-degenerate e$^T$ orbitals directed toward neighboring cobalt atoms of a CoO$_2$ plane[15]. The observation[12] of a transition from polaronic to itinerant electrons in the CoO$_2$ planes as x increases to beyond x = 0.28 shows that in Na$_{0.57}$CoO$_2$, the Co-Co interactions in the CoO2 planes are strong enough to broaden the e$^T$ orbitals into a narrow band of mostly itinerant-electron states, but the large compressibility of these planes in the pressure range 0 < P < 22 kbar is understandable from the virial theorem if strong-correlation fluctuations persist in this band as was deduced previously[13] to account for the T$^{3/2}$ dependence of ρ(T) below 175 K. The observation[7] of a smooth transition from a metallic to a polaronic temperature dependence of the c-axis conductivity in the interval 150 K < T < 250 K is characteristic of a transition from itinerant a$_1^T$ electrons at lowest temperatures to polaronic a$_1^T$ electrons at room temperature, the two phases coexisting in the crossover temperature interval 150 K < T < 250 K. Since pressure increases the c-axis Co-O-O-Co interactions, it follows from the virial theorem that we should expect a large c-axis compressibility at room temperature in the pressure interval where there is a smooth, two-phase transition back to itinerant a$_1^T$ electrons. We therefore identify the large c-axis compressibility in the pressure range 22 < P < 35 kbar with a two-phase transition at room temperature from polaronic back to itinerant behavior of a a$_1^T$ electrons. The existence of a narrow a$_1^T$ band of itinerant-electron states requires unusually strong Co-O-O-Co interactions along the c-axis. How this is possible is made evident by the observation of O$_2$ evolution from the CoO$_2$ sheets on the extraction of more than half of the Na atoms. The evolution of oxygen occurs on oxidation of a redox couple that is pinned at the top of an O$^{2-}$:2p$^6$ band[16]. Pinning of a redox couple occurs where the redox couple lies below the O$^{2-}$:2p$^6$ energy level in a point-charge model; covalent mixing of M-3d and O-2p states lifts the antibonding states to the top of the O:2p$^6$ bands. These antibonding states have the symmetry of the 3d orbitals, but a dramatic increase in the O-2p component of these states occurs where the cationic redox couple falls below the O$^{2-}$:2p$^6$ energy in the point-charge model. In Na$_{1-x}$CoO$_2$, the low-spin Co(IV)/Co(III) redox couple falls below the O-2p$_\pi$ states of the point-charge O$^{2-}$:2p$^6$ energies, and the predominantly O-2p$_\pi$ character of the holes introduced into the pinned antibonding band allows them to combine in surface states to form O$_2^{2-}$ peroxide ions followed by O$_2$ evolution. It is the strong O-2p$_\pi$ character of the ligand-field t$_2$ orbitals of the formal Co(IV)/Co(III) redox couple pinned at the top of the O$^{2-}$:2p$^6$ band that is responsible for the sizeable Co-O-O-Co overlap integral of the a$_1^T$ orbitals.

Fig. 4 is a schematic representation of the antibonding e$^T$ and a$_1^T$ bands pinned at the top of the nominal O$^{2-}$:2p$^6$ bands. Narrowing of the a$_1^T$ band on raising the temperature in the interval 0 < T < 300 K reflects the transition from itinerant to polaronic (or variable-range hopping) behavior of the a$_1^T$ electrons. Placement of the a$_1^T$ band relative to the Fermi energy $E_F$ is deduced from considerations of the thermoelectric power and the thermal-expansion data, and it is consistent with elaborated band structure calculations[17].

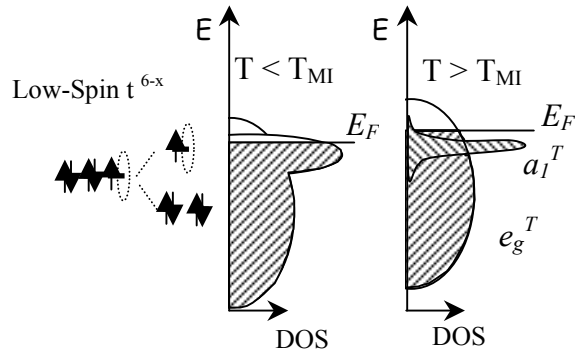

Fig.4: Schematic band diagram of the system $Na_xCoO_2$. The dotted circle represents a hole introduced by $Mn^{4+}$. $T_{MI}$ is the temperature at which the resistivity becomes thermally activated.

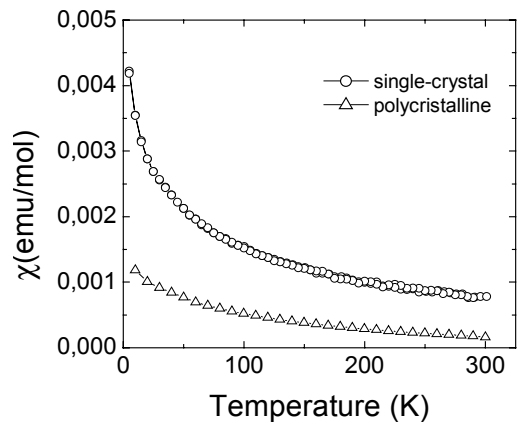

Fig 5: Temperature dependence of the magnetic susceptibility along the ab-plane of a single crystal (circles), and in a polycristalline sample (triangles).

From the virial theorem, the large thermal expansion of the c-axis in the interval 150 K < T < 250 K reflects the change in the $(M-O)_{loc}/(M-O)_{itin}$ ratio as the volume fraction of the polaronic phase among the $a_1^T$ electrons increases with temperature. On the other hand, the $e^T$ orbitals are broadened into a band of itinerant-electron states by Co-Co interactions. The strength of these interactions is also enhanced by the covalent admixture of O-2p character into the Co-3d orbitals. However, the $e^T$ bands are also at the narrow-band limit for itinerant-electron behavior, and a thermal expansion of the Co-Co separation would induce polaronic behavior, which would increase the basal-plane thermal expansion in a positive feedback. Since this situation is not observed, it is apparent that $Na_{0.57}CoO_2$ adjusts its c-axis so as to retain a nearly constant Co-Co distance with increasing temperature. By expanding the c-axis, which induces a transformation to polaronic behavior of the $a_1^T$ holes, electrons can be transferred from the $e^T$ bands to the $a_1^T$ bands so as to retain the Co-Co separation. This interpretation is supported by the almost identical in-plane Co-Co distances in $A_xCoO_{2-\delta}$, $Co^{3+}/Co^{4+} \approx 1$, when A goes from Li to Na to K and even Sr, in spite of the large variation of the Co-Co distance along the c-axis. Therefore, we place the Fermi energy in both the $a_1^T$ and $e^T$ bands, but we place $E_F$ near the top of the $a_1^T$ band in order to account for the thermoelectric power. The general expression for the thermoelectric power as derived from the Boltzmann equation is[18]:

$$\alpha = \frac{1}{e\sigma T}\int (E-E_F)\sigma(E)\frac{\partial f}{\partial E}dE$$

where $f$ is the Fermi-Dirac distribution function and $\sigma(E)$ is defined in

$$\sigma = \int \sigma(E)\frac{\partial f}{\partial E}dE$$

For a normal broad-band metal, Mott[19] assumed that $\sigma(E) \propto E^x$ in the neighborhood of $E_F$ to obtain the expression

$$\alpha \approx \frac{\pi^2 k_B^2 T}{3eE_F}x$$

In our narrow-band case, $E_F$ as measured from the top of the bands is small, and x depends on the curvature of $E_k$ vs $k$. Although the Mott equation (3) may not be applicable to the crossover from itinerant to polaronic behavior, nevertheless the extreme asymmetry of the density of states either side of $E_F$ as pictured in Fig. 4 can be expected to account, according to Eq. (2), for the large value of α(T). The thermoelectric power is sensing the energy dependence of the density of states around $E_F$. From the general expression of equation (2) it can be deduced that when N(E) is an even function around $E_F$, α goes to zero, but increases as the asymmetry in N(E) in the interval $E_F \pm T$ does. From the band diagram of Fig. 4, N(E) changes rapidly with energy near E= $E_F$ which produces a considerable enhancement of α as the $a_1^T$ narrows with temperature. A reduction in the hole doping (increasing the Na content per formula unit close to 0.75) places the $E_F$ further above the $a_1^T$ band, which makes the thermopower smaller (as is observed) in spite of the lower conductivity of the x = 0.7 compound.

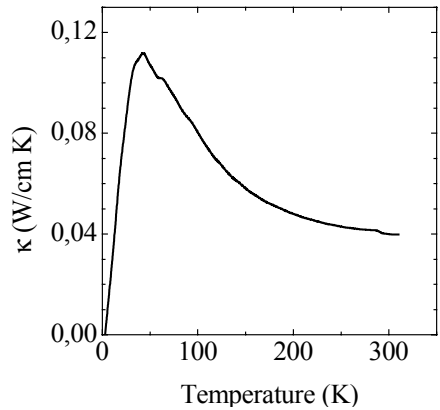

Fig.6: Temperature dependence of the lattice thermal conductivity of a cold-press sintered polycrystalline sample.

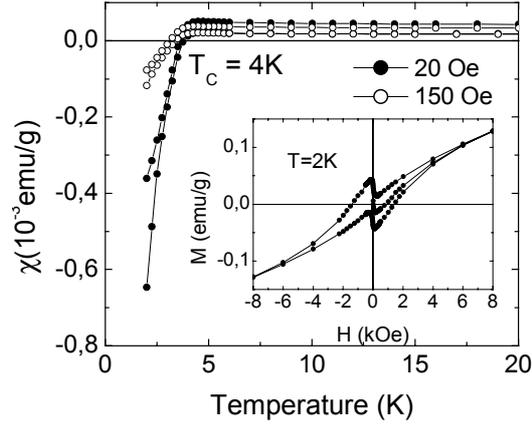

Fig.7: Magnetization curves in FC-ZFC of $Na_xCoO_2 nH_2O$ at 20 Oe and 150 Oe. The $T_C$ is 4 K. Inset: Magnetization vs. field histeresis loop taken at 2 K.

Moreover, $\alpha(T)$ also appears to reflect the electron transfer that was postulated to suppress the thermal expansion of the Co-Co bond in the $CoO_2$ planes. Reduction of the holes in the $a_1^T$ band increases the contribution to $\alpha(T)$ from the $a_1^T$ holes more strongly than it decreases the contribution to $\alpha(T)$ from the $e^T$ holes, so $\alpha(T)$ is enhanced where the $a_1^T$ band is narrowed by the c-axis thermal expansion. Suppression by pressure of the narrowing of the $a_1^T$ band reduces the electron transfer from the $e^T$ to the $a_1^T$ bands and hence the change in slope of $\alpha(T)$ near 100 K.

The coexistence of localized spins and itinerant electrons adds a Curie-Weiss term to the Pauli paramagnetism to give a temperature dependence to the paramagnetic susceptibility of $Na_{0.57}CoO_2$ (Figure 5). This temperature dependence is present in the susceptibility of polycrystalline materials and of the ab-plane of single crystals. The coexistence of strong-correlation fluctuations and itinerant-electron behavior can be expected to introduce bond-length fluctuations that suppress the phonon contribution to the thermal conductivity as has been observed in the $RniO_3$ family of perovskites[20]. In fact, a suppressed thermal conductivity characteristic of an amorphous compound has been reported[9] for nominal $Na0.5CoO_2$. In order to check this result, the thermal conductivity $\kappa(T)$ was measured on a sample prepared from a cold-pressed polycrystalline powder that was resintered at 900°C to grow the grain size. After this treatment, the AAA showed no significant change in the Na/Co ratio. Fig. 6 shows the ambient-pressure $\kappa(T)$ curve. It is evident that a sizeable phonon component is observed with no apparent suppression in the interval 150 K < T < 250 K where the transition from itinerant to polaronic $a_1^T$ electrons has been deduced. Since the random distribution of $Na^+$ ions in the Na planes perturbs the c-axis periodic potential, we introduce Anderson-localized states at the wings of the narrow $a_1^T$ band; the polaronic $a_1^T$ holes may, in fact, be trapped in these states. In this case, the polaronic phase does not fluctuate, and any suppression of the phonon contribution to $\kappa(T)$ in $Na_{0.57}CoO_2$ would come from strong-correlation fluctuations in the $e^T$ bands. But in this compound, the volume fraction of strong-correlation fluctuations in the $e^T$ bands is too small to suppress the phonons completely. Comparison of $\kappa(T)$ for the $RniO_3$ family shows that retention of a significant phonon contribution to $\kappa(T)$ is consistent with a minor volume fraction of strong-correlation fluctuations among the $e^T$ electrons.

Finally, we would like to make a comment about the recently reported superconductivity in Na$_{0.35}$CoO$_2$·1.3H$_2$O[10]. Chemical extraction of Na$^+$ with Br$_2$ and subsequent H$_2$O intercalation between the CoO$_2$ planes produces an increse of the c-axis parameter of ≈ 75% and keeps the in-plane Co-Co distance almost unchanged. It also produces a diamagnetic response and a a drop in the resistivity below *ca.* 4 K . We were able to reproduce the results reported by Takada *et al.*; however, the value of the diamagnetic susceptibility wass lower in our case, signaling a smaller amount of diamagnetic phase in our sample (see figure 7). In fact, the drop to zero resistivity has not been achieved by any group to date. Cold-pressing the sample in order to obtain contact between grains eliminated water and suppressed the evidence for superconductivity. The hysteresis loop at 2 K (inset of figure 7) shows a type II superconductor with a lower critical field of ≈ 125 Oe.

To account for these observations we propose the following tentative model: anhydrous Na$_{0.35}$CoO$_{2-\delta}$ contains oxygen vacancies that perturb the periodic potential, thus suppresing any possibility of superconductivity. The observation that superconductivity is found in nominal Na$_{0.35}$CoO$_2$·1.3H$_2$O, but with such a large amount of water in the Na$^+$ layers, provides a clue to the essential chemistry of the superconductive phase. Oxygen of the intercalated water enters the oxygen vacancies of the CoO$_2$ sheets as occurs for bound water on the surface of an oxide particle. As the highly oxidized, low-spin Co(IV)/Co(III) ions create a strongly acidic CoO$_2$ sheet, the bound water gives its protons to the free interstitial to create (H$_3$O)$^+$ ions, *i.e.*

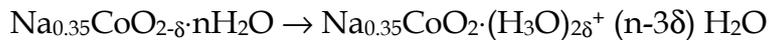

Na$_{0.35}$CoO$_{2-\delta}$·nH$_2$O → Na$_{0.35}$CoO$_2$·(H$_3$O)$_{2\delta}{}^+$ (n-3δ) H$_2$O

provided the amount of interstitial water is large enough that the proton transfer does no make it too acidic to receive all the protons. The ·(H$_3$O)$^+$ ions reduce the CoO$_2$ sheets like the Na$^+$ ions and the bound O$^{2-}$ ions eliminiate the strong perturbation of the periodic potential. If this model is correct, we might expect to find superconductivity in anhydrous Na$_{1-x}$CoO$_2$ at the limiting value of x before oxygen evolution, *i.e.* near x=0.5.

**CONCLUSIONS**

From this study, we draw the following conclusions:

(1) The nominal compositions Na$_{0.5}$CoO$_2$ reported in the literature are probably closer to Na$_{0.6}$CoO$_2$ unless additional Na was extracted chemically or electrochemically. Moreover, these compositions are moisture sensitive, so measured samples may contain an unknown amount of water unless specific precautions are taken to avoid prolonged exposure to humid air.

(2) Oxygen evolution on extraction of more than half the sodium from NaCoO$_2$ signals that the nominal Co(IV)/Co(III) redox couple is pinned at the top of the nominal O$^{2-}$:2p$^6$ bands, which introduces a dominant O-2p$_\pi$ character into the antibonding a$_1{}^T$ and e$^T$ bands of the low-spin cobalt atoms. Consequently, the Co-O-O-Co c-axis interactions and the Co-Co basal-plane interactions are just strong enough to make itinerant at lowest temperatures the $a_1^T$ and $e^T$ holes introduced into these bands by the removal of ≈0.43 sodium atoms in $Na_{0.57}CoO_2$

(3) A large c-axis thermal expansion in the temperature interval 150 K < T < 250 K correlates with a smooth transition from a metallic to a non-metallic temperature dependence of the c-axis resistivity ρ(T); the basal-plane resistivity remains metallic for all temperatures T < 300 K, varying as $T^{3/2}$ below T = 175 K, and the Co-Co separation holds essentially constant throughout the range 77 K < T < 300 K. From the virial theorem, the smooth transition in ρ(T) may be interpreted to reflect a first-order transition with the coexistence of two electronic phases for localized (polaronic) and itinerant $a_1^T$ holes with $(Co-O)_{loc} > (Co-O)_{itin}$. Electron transfer from the $e^T$ band to the overlaping $a_1^T$ band with increasing temperature can account for a temperature-independent Co-Co separation below 300 K.

(4) A first-order transition from a phase with polaronic holes to one with itinerant holes in the $e^T$ bands is found as x increases in $Na_{1-x}CoO_2$, and a large basal-plane room-temperature compressibility in the pressure range P < 22 kbar as well as the $ρ_{ab} \propto T^{3/2}$ dependence below 175 K are consistent with retention of a minority volume of strong-correlation fluctuations in the $e^T$ bands at $Na_{0.57}CoO_2$. An anomalously large room-temperature c-axis compressibility of $Na_{0.57}CoO_2$ in the range 22 < P < 35 kbar may be logically assumed to reflect a smooth transition back to metallic holes in the $a_1^T$ band at room temperature in this pressurerange.

(5) The large thermoelectric power α(T) and its pressure dependence may be interpreted with this model if the top of the narrow $a_1^T$ band is located near $E_F$ so as to give a large curvature of the $E_k$ vs $k$ curves near $E_F$.

(6) The phonon component of the thermal conductivity is not fully suppressed in a cold-pressed, anhydrous ceramic sample with large grains. We infer from this observation that the two $a_1^T$ electronic phases in the interval 150 K < T < 250 K do not fluctuate whereas the minority volume of strongly correlated electrons in the $e^T$ bands may fluctuate.

(7) Trapping of holes in Anderson localized $a_1^T$ states at the top of the band may take place above 150 K as transfer of electrons from the $e^T$ bands to the $a_1^T$ band shifts $E_F$ across a mobility edge.

(8) The observation of superconductivity below 4 K in $Na_{0.35}CoO_2·1.3H_2O$ is incompatible with loss of oxygen from anhydrous $Na_{0.35}CoO_{2-δ}$ unless the reaction $Na_{0.35}CoO_{2-δ}·nH_2O \rightarrow Na_{0.35}CoO_{2-δ}·(H_3O)_{2δ}^+ (n-3δ)H_2O$ occurs.

ACKNOWLEDGEMENTS

We want to acknowledge to Dr. Brian Chushing and Dr. Elin Winkler for helpful discussion and comments. The authors thank the NSF, the Robert A. Welch Foundation, Houston, TX, and the TCSUH of Houston, TX for financial support. F. R. would like to thank the Fulbright Foundation and MECD (Spain) for financial support.